\documentclass[prl,twocolumn,showpacs,superscriptaddress,preprintnumbers,amsmath,amssymb]{revtex4}

\usepackage{graphicx}
\usepackage{dcolumn}
\usepackage{bm}

\begin{document}



\title{Different Evolution of Intrinsic Gap in Kondo Semiconductors SmB$_6$ and YbB$_{12}$} 

\author{J.~Yamaguchi}
 \affiliation{Graduate School of Engineering Science, Osaka University, Toyonaka, Osaka 560-8531, Japan}

\author{A.~Sekiyama}
 \affiliation{Graduate School of Engineering Science, Osaka University, Toyonaka, Osaka 560-8531, Japan}
 \affiliation{SPring-8/RIKEN, Sayo, Hyogo 679-5148, Japan}

\author{M.~Y.~Kimura}
\author{H.~Sugiyama}
\author{Y.~Tomida}
\author{G.~Funabashi}
\author{S.~Komori}
 \affiliation{Graduate School of Engineering Science, Osaka University, Toyonaka, Osaka 560-8531, Japan}

\author{T.~Balashov}
\author{W.~Wulfhekel}
 \affiliation{Physikalisches Institut, Karlsruhe Institute of Technology (KIT), 76131 Karlsruhe, Germany}

\author{T.~Ito}
\author{S.~Kimura}
 \affiliation{UVSOR Facility, Institute for Molecular Science, Okazaki 444-8585, Japan}

\author{A.~Higashiya}
\author{K.~Tamasaku}
 \affiliation{SPring-8/RIKEN, Sayo, Hyogo 679-5148, Japan}

\author{M.~Yabashi}
\author{T.~Ishikawa}
 \affiliation{SPring-8/RIKEN, Sayo, Hyogo 679-5148, Japan}
 \affiliation{SPring-8/JASRI, Sayo, Hyogo 679-5198, Japan}

\author{S.~Yeo}
 \affiliation{Korea Atomic Energy Research Institute, Daejeon 305-600, Republic of Korea}
 
\author{S.-I.~Lee}
 \affiliation{Department of Physics, Sogang University, Seoul 121-742, Republic of Korea}

\author{F.~Iga}
\author{T.~Takabatake}
 \affiliation{Graduate School of Advanced Sciences of Matter, Hiroshima University, Higashi-Hiroshima 739-8526, Japan}

\author{S.~Suga}
 \affiliation{Graduate School of Engineering Science, Osaka University, Toyonaka, Osaka 560-8531, Japan}
 \affiliation{SPring-8/RIKEN, Sayo, Hyogo 679-5148, Japan}

%


\date{\today}

\begin{abstract}
Dependence of the spectral functions on temperature and rare-earth substitution was examined in detail for Kondo semiconductor alloys Sm$_{1-x}$Eu$_x$B$_6$ and Yb$_{1-x}$Lu$_x$B$_{12}$ by bulk-sensitive photoemission. It is found that the $4f$ lattice coherence and intrinsic (small) energy gap are robust for SmB$_6$ against the Eu substitution up to $x$ = 0.15 while both collapse by Lu substitution already at $x$ = 0.125 for YbB$_{12}$. Our results suggest that the mechanism of the intrinsic gap formation is different between SmB$_6$ and YbB$_{12}$ although they were so far categorized in the same kind of Kondo semiconductors.
\end{abstract}

\pacs{75.20.Hr, 75.30.Mb, 79.60.-i}

\maketitle

Strongly correlated electron systems based on rare-earth (RE) elements have attracted wide interest in the past few decades due to their intriguing properties, such as valence-fluctuation (VF), metal-insulator (-semiconductor) transitions, and heavy-fermion superconductivity near the quantum critical point. Among them, SmB$_6$ \cite{PRL22_295} and YbB$_{12}$ \cite{JMMM31-34_437and47-48_429} are well known as typical VF Kondo semiconductors. At high temperatures the Kondo semiconductors behave as metals with localized $f$ magnetic moments, whereas a narrow gap opens at the Fermi level ($E_F$) below the characteristic temperature $T^*$, which has been reported as $T^* \sim$ 140 K for SmB$_6$ \cite{JPCS63_1223} and $T^* \sim$ 80 K for YbB$_{12}$ \cite{JMMM177-181_337}. 
The gaps of single crystal SmB$_6$ and YbB$_{12}$ have been deduced from the transport, optical, and inelastic neutron scattering (INS) measurements. The so-called hybridization-gap model has been proposed to explain the gap formation. Recently, two types of gaps with different magnitude were reported as the so-called ``large'' gap ($\Delta_{\rm L} \sim$ 10--20 meV) and ``small'' gap ($\Delta_{\rm S} \sim$ 3--7 meV) \cite{gaps}, where the latter takes place for the in-gap states formed within the large gap. The in-gap states are often seen in Kondo semiconductors (not only SmB$_6$ \cite{PRB59_1808} and YbB$_{12}$ \cite{PRB73_045207} but also Ce$_3$Bi$_4$Pt$_3$ \cite{PRB77_115134} and FeSi \cite{PRL71_1748}). 
Various theoretical approaches have been performed to interpret the evolution of the large and small gaps as well as the VF behavior by means of such as Wigner lattice model \cite{JPC37_C4}, exciton-polaron model \cite{JPCM7_307}, and Anderson lattice model at half filling \cite{PRB68_235213}. 
Despite numerous experimental and theoretical studies, the origin of the gap states is not yet clarified.

Photoelectron spectroscopy (PES) is one of the powerful tools to directly determine the density of state (DOS) as well as quasiparticle band structures. To date, PES studies, using conventional photon energies ($h\nu \sim$ 20--100 eV), have been performed for SmB$_6$ \cite{JPCS63_1223} and YbB$_{12}$ \cite{PRL77_4269and82_992}. These conventional PESs have, however, the inherent problem of the surface-sensitivity due to the short inelastic mean-free path (IMFP) of photoelectrons \cite{SIA35_268}, where the IMFP is restricted to less than $\sim$5 \AA. Since the surface of such Kondo semiconductors are thought to be metallic \cite{PhysicaB293_417}, more bulk-sensitive PES studies are desired. In this decade, hard x-ray PES (HAXPES) has been widely recognized as a highly bulk-sensitive technique because of the long IMFP and intensively applied to various strongly correlated electron systems \cite{HAXPES}. The IMFP reaches up to $\sim$100 \AA\ at $h\nu \sim$ 8 keV. Recently, extremely low-energy ($h\nu \leqslant$ 10 eV) PES (ELEPES) with excitations by laser, synchrotron radiation (SR), or Xe resonance line, has also become popular as a bulk-sensitive ultra-high-resolution technique \cite{ELEPES}. The IMFP for ELEPES is expected to reach values comparable to that for HAXPES under certain conditions. Thus, combined HAXPES and ELEPES studies are extremely useful for revealing intrinsic bulk electronic structures for SmB$_6$ and YbB$_{12}$.

In our recent HAXPES experiments for the Kondo semiconductor alloys Yb$_{1-x}$Lu$_x$B$_{12}$ \cite{PRB79_125121}, it is concluded that the Yb $4f$ lattice coherence is effective for $x$ = 0, which collapses easily by Lu substitution (already at $x$ = 0.125). In addition, optical studies have independently shown that the gap collapses for $x$ = 0.125 \cite{PRB62_R13265}. From these results, it has been deduced that the Yb $4f$ lattice coherence plays an essential role for the gap formation in pure YbB$_{12}$. In the INS studies \cite{JPCM16_2631}, however, it was suggested that the spin-gap of $\sim$10 meV could be driven by the Yb $4f$ single-site effects because the gap can be robust up to $x$ = 0.9. To fully understand the origin of the gaps in Kondo semiconductors, the competition between $4f$ lattice coherence effects and $4f$ single-site effects must be carefully studied in various cases. 
Here, we report on bulk-sensitive HAXPES and ELEPES on the Kondo semiconductor alloys Sm$_{1-x}$Eu$_x$B$_6$ and Yb$_{1-x}$Lu$_x$B$_{12}$ including the doping and temperature dependence.

Single crystals of Sm$_{1-x}$Eu$_x$B$_6$ (cubic CaB$_6$ type: $x$ = 0, 0.15, and 0.5) \cite{APL94_042509} and Yb$_{1-x}$Lu$_x$B$_{12}$ (cubic UB$_{12}$ type: $x$ = 0 and 0.125) \cite{JMMM177-181_337} were grown by Al flux and floating-zone methods. HAXPES measurements were performed with SR ($h\nu \sim$ 8 keV) at BL19LXU of SPring-8 \cite{PRL87_140801}, using an MBS A1-HE spectrometer. ELEPES measurements were performed with SR ($h\nu$ = 7 eV) and an MBS A1 spectrometer at BL7U of UVSOR-II \cite{AIP879_527}, as well as with an RF-excited MBS T-1 lamp (Xe I resonance line: $h\nu$ = 8.4 eV) and a SCIENTA SES2002 spectrometer in Osaka University \cite{JJAP47_2265}. 
Clean surfaces of all samples were obtained by ${\it in\ situ}$ fracturing in a vacuum with a base pressure of $\sim$$5\times10^{-8}$ Pa. The energy calibration was performed with the Au Fermi edge at each temperature. The total-energy resolution was set to 120 meV for Sm$_{1-x}$Eu$_x$B$_6$ and 65 meV for Yb$_{1-x}$Lu$_x$B$_{12}$ \cite{PRB79_125121} in HAXPES and 6 meV for both systems in ELEPES.

\begin{figure}
\includegraphics[width=7.5cm,clip]{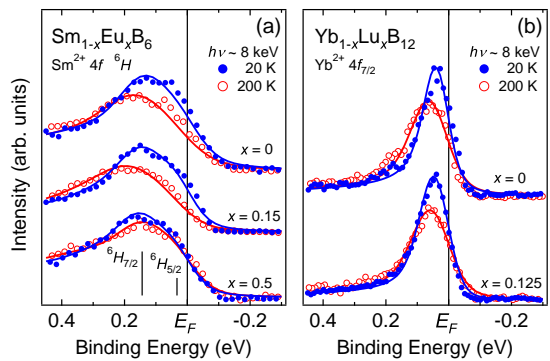}
\begin{center}
\caption{(color online). HAXPES spectra at 20 and 200 K for (a) Sm$_{1-x}$Eu$_x$B$_6$ and (b) Yb$_{1-x}$Lu$_x$B$_{12}$ \cite{PRB79_125121}. The circles and solid lines indicate the experimental data and the fitting results (see text). The spectra are normalized for different $x$ by the integrated intensity in the binding energy range of $-$0.2 eV $\leqslant E_B \leqslant$ 0.6 eV for Sm$_{1-x}$Eu$_x$B$_6$ and $-$0.2 eV $\leqslant E_B \leqslant$ 0.8 eV for Yb$_{1-x}$Lu$_x$B$_{12}$ at each temperature to compare the peak positions.} 
\label{fig:fig1} 
\end{center}
\end{figure}

Figure~\ref{fig:fig1}(a) shows the doping dependence of the HAXPES spectra at 20 and 200 K for Sm$_{1-x}$Eu$_x$B$_6$. For comparison, results of Yb$_{1-x}$Lu$_x$B$_{12}$ are also reproduced \cite{PRB79_125121} in Fig.~\ref{fig:fig1}(b). 
At $h\nu \sim$ 8 keV the photoionization cross sections \cite{NDT32_1} of Sm and Yb $4f$ states are higher than those of B $2sp$ and RE $5d$ states. Therefore, the spectra near $E_F$ at $h\nu \sim$ 8 keV for these materials are dominated by the excitations $f^6\rightarrow f^5$ for Sm$^{2+}$ ($4f^6$) and $f^{14}\rightarrow f^{13}$ for Yb$^{2+}$ ($4f^{14}$) states \cite{EuandLu}. 
The broad peaks for Sm$_{1-x}$Eu$_x$B$_6$ consist of the Sm $4f^5$ ($^6H_{5/2}$ and $^6H_{7/2}$) final-state multiplets \cite{SB23_59} [line spectra in Fig.~\ref{fig:fig1}(a)], whereas the single peaks for Yb$_{1-x}$Lu$_x$B$_{12}$ are ascribed to the Yb $4f^{13}_{7/2}$ final-state. In Fig.~\ref{fig:fig1}(a), the Sm $4f^5$ peaks for $x$ = 0 and 0.15 in Sm$_{1-x}$Eu$_x$B$_6$ clearly show the energy shift toward $E_F$ on going from 200 to 20 K, while those for $x$ = 0.5 no longer show such an energy shift. To estimate the value of the peak shift, the spectra were fitted with Lorentzian (for the slight life time broadening) and Gaussian (for the instrumental resolution) broadened functions as shown by the solid lines in Fig.~\ref{fig:fig1}(a). The relative intensity and energy for the Sm $4f$ multiplets are properly taken into account \cite{SB23_59}. The estimated peak shift toward $E_F$ on cooling for $x$ = 0 and 0.15 is $\sim$40 meV, while that for $x$ = 0.5 is less than 10 meV in Sm$_{1-x}$Eu$_x$B$_6$. On the other hand, the Yb $4f$ peak shift toward $E_F$ is $\sim$20 meV for $x$ = 0 whereas it is at most 10 meV for $x$ = 0.125 in Yb$_{1-x}$Lu$_x$B$_{12}$ [Fig.~\ref{fig:fig1}(b)]. The temperature dependence of the $4f$ peaks observed for $x$ = 0 and 0.15 in Sm$_{1-x}$Eu$_x$B$_6$ is qualitatively similar to that for pure YbB$_{12}$, in which the Yb $4f$ lattice coherence is essential for both peak shift and gap formation. Such a $4f$ peak shift with temperature, derived from the $4f$ lattice coherence effects, has also been discussed in YbAl$_3$ \cite{YbAl3} and YbInCu$_4$ \cite{YbInCu4}. In this sense, the Sm $4f$ lattice coherence survives not only for $x$ = 0 but also for $x$ = 0.15 in Sm$_{1-x}$Eu$_x$B$_6$.

\begin{figure}
\begin{center}
\includegraphics[width=8cm,clip]{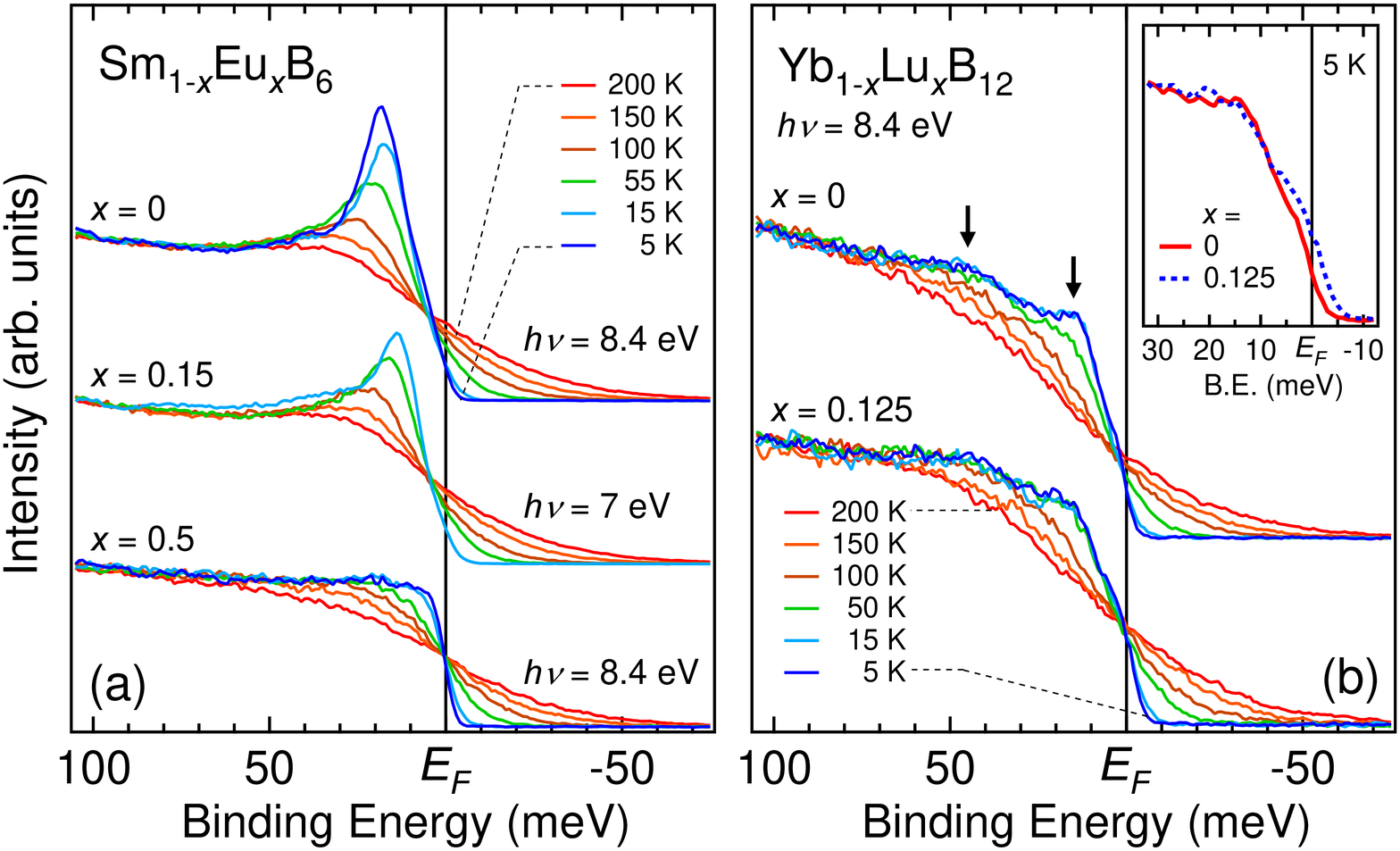}
\caption{(color online). Temperature dependent ELEPES spectra for (a) Sm$_{1-x}$Eu$_x$B$_6$ and (b) Yb$_{1-x}$Lu$_x$B$_{12}$. The inset of (b) describes the doping dependence of the spectra for Yb$_{1-x}$Lu$_x$B$_{12}$ at 5 K. The intensity of all these spectra is normalized at $E_B$ = 100--120 meV, where the spectra show no temperature dependence in both systems.}
\label{fig:fig2}
\end{center} 
\end{figure}

In order to study the details of the relation between the gap formation and the $4f$ lattice coherence for both systems overcoming the resolution limit of HAXPES, we have performed the ELEPES with high resolution. The doping and temperature dependent ELEPES spectra near $E_F$ for Sm$_{1-x}$Eu$_x$B$_6$ and Yb$_{1-x}$Lu$_x$B$_{12}$ are presented in Figs.~\ref{fig:fig2}(a)~and \ref{fig:fig2}(b). The observed spectra are dominated by the non-$4f$ states, i.e., B $2sp$ and RE $5d$ states hybridized with the Sm or Yb $4f$ states \cite{NDT32_1}. 
In SmB$_6$, the spectral weight at $E_F$ decreases gradually and the peak narrows and shifts toward $E_F$ with decreasing the temperature from 200 to 5 K. The so-called leading-edge of the spectra is clearly observed on the occupied state side, demonstrating a finite gap at low temperatures. The spectra show a prominent peak at 18 meV below 15 K. 
The spectra of Sm$_{1-x}$Eu$_x$B$_6$ with $x$ = 0.15 show the similar temperature dependence while the spectra for $x$ = 0.5 show a typical metallic thermal behavior even at low temperatures without a prominent peak. 
This narrow peak at low temperatures has also been observed at the same binding energy in several spectra measured at $h\nu$ between 7 and 12 eV for $x$ = 0.15 (not shown here). 
Therefore, both matrix element effects and surface contributions can be neglected.
In transport measurements, an exponential increase in the electrical resistivity due to the gap formation on cooling has been observed for $x \leqslant$ 0.2, in consistence with the present ELEPES results. 
%
%

In the case of YbB$_{12}$, the spectra show a slight decrease of the intensity at $E_F$ upon cooling, suggesting a gap formation [Fig.~\ref{fig:fig2}(b)], where two peaks are observed at 15 and 45 meV at low temperatures (indicated by arrows) like the previous work by SR at $h\nu =$ 15.8 eV \cite{PRB73_033202}. For the Lu doping up to $x$ = 0.125, the two peaks are located at almost the same positions. However, a clear difference between for $x$ = 0 and 0.125 is seen at 5 K as shown in the inset of Fig.~\ref{fig:fig2}(b). The intensity around $E_F$ is definitely higher for $x$ = 0.125.

Another feature to be noticed in the ELEPES spectra is a different temperature dependence between SmB$_6$ and YbB$_{12}$ in the range of 5--15 K, which are far below the characteristic temperature $T^*$. Namely, the spectral weight of the narrow peak is further enhanced from 15 to 5 K for SmB$_6$, while such a temperature dependence is not seen for YbB$_{12}$. These results have revealed that the electronic state of YbB$_{12}$ is almost in the ground state at 15 K ($T^* \sim$ 80 K) but that of SmB$_6$ is still not in the ground state even at 15 K ($T^* \sim$ 140 K).

\begin{figure*}
\begin{center}
\includegraphics[width=10cm,clip]{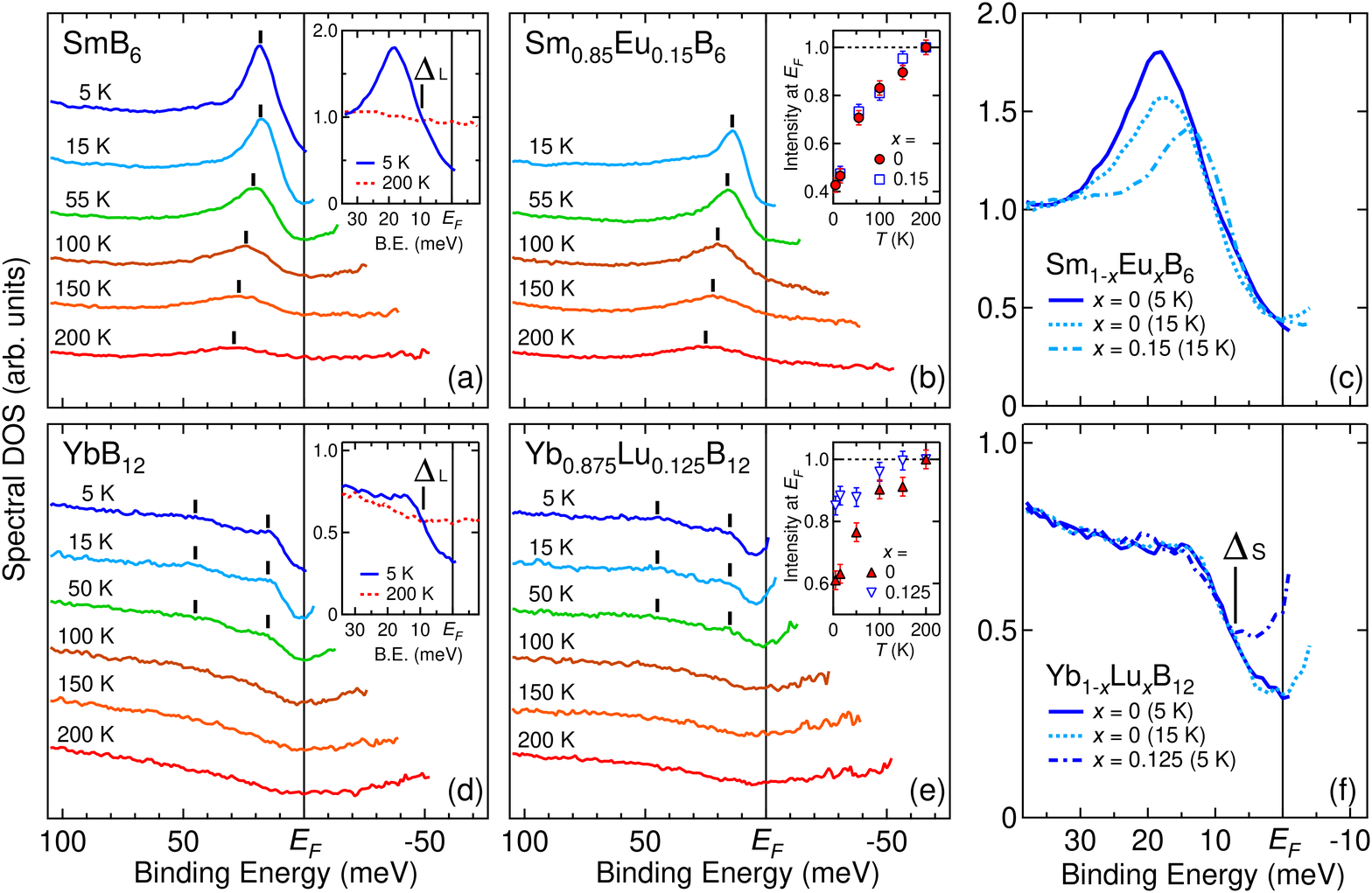}
\caption{(color online). Temperature dependent spectral DOS of Sm$_{1-x}$Eu$_x$B$_6$ [for (a) $x$ = 0 and (b) $x$ = 0.15] and Yb$_{1-x}$Lu$_x$B$_{12}$ [for (d) $x$ = 0 and (e) $x$ = 0.125]. The inset of (a) and (d) shows the comparison of the spectral DOS between 5 and 200 K for SmB$_6$ and YbB$_{12}$. The inset of (b) and (e) is the temperature dependence of the spectral intensity at $E_F$ for Sm$_{1-x}$Eu$_x$B$_6$ and Yb$_{1-x}$Lu$_x$B$_{12}$, where the value of the intensity at each temperature is normalized by that at 200 K. The doping dependence of the spectral DOS at low temperatures for (c) Sm$_{1-x}$Eu$_x$B$_6$ and (f) Yb$_{1-x}$Lu$_x$B$_{12}$.}
\label{fig:fig3} 
\end{center}
\end{figure*}

For advanced discussions on the doping and temperature dependence of the gap behaviors and peak structures, the spectral DOS information has been obtained by dividing the ELEPES spectra by the Fermi-Dirac function convoluted with the instrumental resolution (6 meV). The results are displayed in Fig. \ref{fig:fig3}, where we show the spectral DOS up to $3k_BT$ above $E_F$ keeping high enough statistics. 
The gap for SmB$_6$ is estimated to be $\sim$10 meV from the energy difference between $E_F$ and the intersection point of two spectra at 5 and 200 K [$\Delta_{\rm L}$ in the inset of Fig.~\ref{fig:fig3}(a)]. This energy may correspond to that of the ``large'' gap obtained by other experiments \cite{PhysicaB293_417}. 
In the spectral DOS at 5 K we cannot notice any additional structure such as in-gap state, though the data are of much higher statistics than results so far reported. However, the spectral intensity at $E_F$ is gradually suppressed from $T^* \sim$ 140 K to low temperatures for both $x$ = 0 and 0.15 as shown together in the inset of Fig.~\ref{fig:fig3}(b). The temperature dependence for $x$ = 0.15 is essentially equivalent with that for $x$ = 0, whereas the intensity stays constant for $x$ = 0.5 (data not shown). 
Therefore, one can safely conclude that not only the large gap but also the small gap do not collapse in $x$ = 0.15 Sm$_{1-x}$Eu$_x$B$_6$ up to 50 K. The non observation of the possible in-gap state at $\sim$3 meV below 13 K \cite{PRB59_1808} in our ELEPES may be due to the limited resolution or too low DOS of this state. 
Meanwhile, the peak shifts almost linearly toward $E_F$ with decreasing temperature for both $x$ = 0 and 0.15 as shown in Figs.~\ref{fig:fig3}(a) and~\ref{fig:fig3}(b). It is, however, found that the peak energy for $x$ = 0.15 is $\sim$5 meV smaller than $x$ = 0 at each temperature. Furthermore, the peak intensity is lower for $x$ = 0.15 [see a comparison of the spectral DOS between $x$ = 0 and 0.15 at 15 K in Fig.~\ref{fig:fig3}(c)]. In our Sm $3d$ HAXPES studies of Sm$_{1-x}$Eu$_x$B$_6$ \cite{SmEuB6HAXPES}, the estimated bulk Sm valence increases from $\sim$2.58 for $x$ = 0 to $\sim$2.68 for $x$ = 0.15 at 20 K, suggesting the decrease of the hybridization between the Sm $4f$ and non-$4f$ (B $2sp$ and RE $5d$) states with rising $x$. Thus the peak shift and the decrease in the peak intensity with $x$ observed in the ELEPES can be understood by considering the reduction of the $4f$ hybridization.

In Yb$_{1-x}$Lu$_x$B$_{12}$ the spectral DOS for $x$ = 0 and 0.125 is shown in Figs.~\ref{fig:fig3}(d) and \ref{fig:fig3}(e). For both, the intensity of the peaks at 15 and 45 meV (the origin of these two peaks is discussed in Refs.~\cite{PRB73_033202,PRL94_247204and99_137204}) reduces with increasing temperature and the peaks are almost smeared out above $\sim$100 K. 
One can see no remarkable changes of the temperature dependence of the peaks by Lu substitution. The gap of YbB$_{12}$ is estimated to be $\sim$10 meV [see the inset of Fig.~\ref{fig:fig3}(d)], in consistence with the ``large'' gap reported by the electrical resistivity \cite{JMMM177-181_337}. As seen in Figs.~\ref{fig:fig3}(d) and~\ref{fig:fig2}(b), the intensity at $E_F$ for $x$ = 0 decreases upon cooling as indicated by red triangles ($\blacktriangle$) in the inset of Fig.~\ref{fig:fig3}(e). Below $T^* \sim$ 80 K, it starts to reduce steeply. However, the intensities at $E_F$ for $x$ = 0.125 do not decrease so much as shown in Fig.~\ref{fig:fig3}(e) and the inset [blue triangles ($\bigtriangledown$)]. The direct comparison of DOS between $x$ = 0 and 0.125 at low temperatures is shown in Fig.~\ref{fig:fig3}(f). One finds that the spectral DOS near $E_F$ recovers noticeably for $x$ = 0.125. This energy scale below $\sim$7 meV is comparable to the magnitude of the ``small'' gap \cite{JMMM177-181_337}. We conclude that the small gap ($\Delta_{\rm S} \lesssim$ 7 meV) closes at low temperatures, whereas the large gap ($\Delta_{\rm L} \gtrsim$ 10 meV) still remains with Lu substitution. From the doping dependence of these two gaps for Yb$_{1-x}$Lu$_x$B$_{12}$, we conclude as the small gap collapses when the $4f$ lattice coherence is broken, while the large gap survives in this systems because it is induced by the $4f$ single-site effects. The spin-gap, robustly detected up to $x$ = 0.9 in the INS for Yb$_{1-x}$Lu$_x$B$_{12}$ \cite{JPCM16_2631}, might have such a character as the presently observed large gap.

As already mentioned, the small gap as well as the large gap survive at low temperatures at least up to $x$ = 0.15 and up to 50 K in Sm$_{1-x}$Eu$_x$B$_6$ whereas the small gap collapses already for $x$ = 0.125 even at 5 K in Yb$_{1-x}$Lu$_x$B$_{12}$, where the small gap is responsible for the semiconducting behavior. These results are found to be correlated well with the temperature dependence of the $4f$ spectra in Fig.~\ref{fig:fig1}, which suggests that the $4f$ lattice coherence for SmB$_6$ is more robust than YbB$_{12}$ against the RE substitution. 
Within the periodic Anderson model and/or hybridization-gap model \cite{PRB68_235213,JPSJ67_3151}, the electronic structure of Kondo semiconductors is equivalent to that of usual band insulators at low temperatures due to the renormalization. Therefore, it is expected from these models that the intrinsic gap should be closed at $E_F$ for Kondo semiconductors by the collapse of the $4f$ lattice coherence and/or by the deviation of the electron number from even integers. In addition, these models suggest that the spectral function is independent of temperature in the region well below $T^*$. 
These features are consistent with the results for Yb$_{1-x}$Lu$_x$B$_{12}$ at $T \leqslant$ 15 K, but contradicting the results in Sm$_{1-x}$Eu$_x$B$_6$. One might think as the doping dependence of the $4f$ mean valence could be responsible for the qualitative difference in spectral behaviors between SmB$_6$ and YbB$_{12}$. However, the $4f$ mean valence is clearly enhanced with $x$ for Sm$_{1-x}$Eu$_x$B$_6$ \cite{SmEuB6HAXPES} whereas it is hardly changed in Yb$_{1-x}$Lu$_x$B$_{12}$ on going from $x$ = 0 to 0.125 \cite{PRB79_125121}. Then the mean valence and its doping dependence are found to be rather independent of the collapse of the intrinsic small gap. Our results have revealed that the mechanism of the small gap formation for SmB$_6$ is different from that for YbB$_{12}$, although they have so far been categorized in the same kind of Kondo semiconductors.

In conclusion, combination of bulk-sensitive HAXPES and ELEPES is found to be a very powerful method for studying electronic structures of strongly correlated electron systems as the Kondo semiconductors. It is found that the $4f$ lattice periodicity is essential for the gap formation and collapse in YbB$_{12}$ while such a scenario cannot explain the stability of the small gap against Eu substitution up to $x$ = 0.15 for Sm$_{1-x}$Eu$_x$B$_6$.

This work was supported by the Grant-in-Aid for Science Research (18104007, 18684015, 21340101, and Innovative Areas ``Heavy Electrons" 20102003) of MEXT, Japan, the Global COE program (G10) of JSPS, Japan, and the KHYS, KIT, Germany. This work at Sogang University was supported by MIST/KRF (2009-0051705) of Korea. A part of the ELEPES study was performed as a Joint Studies Program of the Institute for Molecular Science (2008).


\begin{thebibliography}{00}
\bibitem{EuandLu}The $4f^6$ multiplets from Eu$^{2+}$ ($4f^7$) state for Sm$_{1-x}$Eu$_x$B$_6$ ($x$ = 0.15 and 0.5) and the $4f^{13}$ doublets from Lu$^{3+}$ ($4f^{14}$) state for Yb$_{1-x}$Lu$_x$B$_{12}$ ($x$ = 0.125) are observed as the localized states in 1--2 eV and 7--9 eV, and they do not contribute directly to the spectral changes near $E_F$.
\end{thebibliography}
\end{document}